\documentclass[acmsmall]{acmart}
\usepackage{flushend}
\usepackage{
float}
\usepackage{subfigure}
\AtBeginDocument{%
  \providecommand\BibTeX{{%
    \normalfont B\kern-0.5em{\scshape i\kern-0.25em b}\kern-0.8em\TeX}}}
\usepackage{tabularx}
\usepackage{bbding}
\usepackage{graphicx}
\usepackage{caption}
\usepackage{geometry}
\usepackage{amsmath}
\usepackage{makecell}
\usepackage{float}
\usepackage{color}
\usepackage{booktabs}
\usepackage[normalem]{ulem}
\usepackage{tabu}
\usepackage{hyperref}
\usepackage{xcolor}
\usepackage{appendix}

\setcopyright{acmlicensed} \acmJournal{PACMHCI} \acmYear{2025} \acmVolume{9} \acmNumber{2} \acmArticle{CSCW031} \acmMonth{4}\acmDOI{10.1145/3710929} 

\begin{document}

\title[An Image of Ourselves in Our Minds]{"An Image of Ourselves in Our Minds": How College-educated Online Dating Users Construct Profiles for Effective Self Presentation}

\author{Fan Zhang}
\email{zfan1218@gmail.com}
\orcid{0009-0009-9990-8820}
\affiliation{
\institution{City University of Hong Kong}
\city{Hong Kong SAR}
\country{China}}

\author{Yun Chen}
\email{10030362@network.rca.ac.uk}
\orcid{0009-0001-3783-2199}
\affiliation{
\institution{Royal College of Art}
\city{London}
\country{United Kingdom}}

\author{Xiaoke Zeng}
\email{zxioke@outlook.com}
\orcid{0009-0007-5987-502X}
\affiliation{
\institution{City University of Hong Kong}
\city{Hong Kong SAR}
\country{China}}

\author{Tianqi Wang}
\email{tttttiq7@gmail.com}
\orcid{0009-0002-4634-2092}
\affiliation{
\institution{ Guangdong University of Technology}
\city{Guangzhou}
\country{China}}

\author{Long Ling}
\email{lucyling0224@gmail.com}
\orcid{0009-0001-2635-788X}
\affiliation{
\institution{Tongji University}
\city{Shanghai}
\country{China}}

\author{RAY LC}
\authornote{Correspondences should be addressed to LC@raylc.org}
\email{LC@raylc.org}
\orcid{0000-0001-7310-8790}
\affiliation{
\institution{City University of Hong Kong}
\city{Hong Kong SAR}
\country{China}}


\renewcommand{\shortauthors}{Fan Zhang et al.}

\begin{abstract}
Online dating is frequently used by individuals looking for potential relationships and intimate connections. Central to dating apps is the creation and refinement of a dating profile, which represents the way individuals desire to present themselves to potential mates, while hiding information they do not care to share. To investigate the way frequent users of dating apps construct their online profiles and perceive the effectiveness of strategies taken in making profiles, we conducted semi-structured interviews with 20 experienced users who are Chinese college-educated young adults and uncovered the processes and rationales by which they make profiles for online dating, particularly in selecting images for inclusion. We found that participants used idealized photos that exaggerated their positive personality traits, sometimes traits that they do not possess but perceive others to desire, and sometimes even traits they wish they had possessed. Users also strategically used photos that show personality and habits without showing themselves, and often hid certain identifying information to reduce privacy risks. This analysis signals potential factors that are key in building online dating profiles, providing design implications for systems that limit the use of inaccurate information while still promoting self-expression in relationship platforms.
\end{abstract}

\ccsdesc[500]{Human-centered computing~Empirical studies in collaborative and social computing}

\keywords{Profile, Profile Picture, Online Dating, Self-presentation, Authenticity}

\received{January 2024}
\received[revised]{July 2024}
\received[accepted]{October 2024}

\maketitle

\section{Introduction}\label{sec:Introduction}
Online dating has emerged as a pivotal tool for individuals seeking potential relationships and intimate connections. In the digital age, dating apps have revolutionized the way individuals engage in the pursuit of companionship and love \cite{li_female_2022, Shen_seeking_2024}. Central to these applications is the creation and refinement of a dating profile, a critical aspect that dictates how individuals present themselves to potential mates. This process of self-presentation is not just about showcasing oneself but also involves strategic decisions about what to reveal and what to conceal \cite{toma_separating_2008}. The choices made in profile creation, from the selection of photos to the crafting of personal descriptions, play a significant role in shaping the online dating experience \cite{lemke_prevalence_2018, Kane_Nonverbal_2009}.

Studies have shown that self-presentation in online dating profiles is influenced by several factors, including personality traits \cite{Jimenez_Personality_2014}, gender differences \cite{kisilevich_exploring_2011}, and the desire to manage impressions and maintain privacy \cite{merunkova_goffmans_2019,lo_contradictory_2013}. For instance, users may choose to present themselves in a certain way to attract attention, empower themselves, or seek self-verification \cite{lemke_prevalence_2018}. Additionally, gender and cultural differences significantly impact how individuals construct their online personas, with varying patterns of self-disclosure observed across different countries \cite{kisilevich_exploring_2011}. Furthermore, the relationship status of users can affect their self-presentation strategies on social networking sites, with single users often disclosing more personal information and photographs compared to those in relationships \cite{winter_digital_2011}.

Despite the increasing prevalence of online dating, there is a relative lack of understanding of how users construct and optimize their online dating profiles. Existing studies have focused on the authenticity of self-presentation in online dating profiles \cite{toma_separating_2008}. Yet, there is still a lack of understanding of how users choose text and images for effective self-presentation. Furthermore, we know little about how information concealment or embellishment in online dating profiles affects the perception of these profiles. This gap in research is significant, considering the impact of self-presentation may have on online dating experiences and outcomes \cite{peng_consequences_2022}.

In this paper, we define “strategy” as the deliberate and conscious methods individuals use to manage their self-presentation on online dating platforms. These methods primarily involve selecting and organizing photos and text to shape how potential partners perceive them. These strategies include decisions about which traits to highlight or conceal, and how to balance self-expression to optimize their profiles to attract potential partners.

Our research is guided by the following questions:

\textit{\textbf{RQ1:} What strategies do people adopt when selecting images for their dating profiles?}

\textit{\textbf{RQ2:} How do people use dating profiles to present themselves and for what purpose?}

\textit{\textbf{RQ3:} What key factors may influence people's perception of others' profiles?}

To explore these questions, we conducted semi-structured interviews with 20 college-educated young adults in China who are experienced users of dating apps. Our participants were selected to provide diverse insights into the profile creation process. Through these interviews, we uncovered the nuanced processes and rationales behind users' choices when constructing their online dating profiles, especially in photo selection. Our findings reveal that participants often use unrealistic photos that exaggerate positive personality traits, sometimes portraying traits they do not possess but believe others find desirable, and occasionally traits they wish they could have. Additionally, users strategically use photos that display personality and habits without fully revealing their identity, often concealing certain identifying information to mitigate privacy risks.

Our work illustrates how a profile creation process can reveal online dating app users' psychological and social needs. It can contribute to (1) signaling potential key factors in building online dating profiles and (2) showing implications for the design of online dating platforms that allow for authentic self-presentation with the desire for privacy, more precise matchmaking tools for better match outcomes, and establishing common ground for enhanced trust.

\section{Background}\label{sec:Background}

\subsection{Self-presentation in Online Dating}

With rapid technological advancements, online dating has become a mainstream form of social interaction \cite{jia_when_2018}. 
This trend requires users to effectively enhance their attractiveness in self-presentation, because less attractive profiles are frequently quickly neglected among abundant profiles to look through in the online environment \cite{heino_relationshopping_2010,toma_separating_2008,sobieraj_tinder_2022}. As first outlined by Goffman’s impression management theory \cite{goffman1959presentation}, self-presentation is integral to how individuals shape their identities during social interactions within relationships. In the early phases of developing relationships, the way individuals present themselves is crucial, particularly as others need this information to determine whether they want to pursue a relationship \cite{taylor1987communication, Derlega1987SelfdisclosureAR,schlenker1984identities}.

Due to the need to present a desirable and attractive self and compare with others, online daters utilize different strategies to market their “best” selves \cite{toma_separating_2008}, sometimes even exaggerating or “lying” on their profiles. For example, many dating app users emphasize the importance of small cues (e.g., grammar or any given message) in the reduced-cue environment, carefully designing their profiles to attract potential partners \cite{ellison_managing_2006,ellison_profile_2012}. 
These strategies involve carefully selecting photos and writing texts that reflect their personal character, interests, and lifestyle \cite{lloyd_identity_nodate,fiore_homophily_2005,saltes_disability_2013}. Moreover, researchers found that it is ubiquitous but subtle to misrepresent or deceive on physical attributes, as well as other information (e.g., photographs) when creating dating profiles \cite{toma_separating_2008,hancock_putting_2009,hancock_truth_2007}. 
The time-shifted nature of online profiles gives online daters chances to represent their present self using some aspects of their past self and an ideal version of their future selves \cite{ellison_profile_2012}. Some daters may also lie in some descriptors to fit into a broader range of search parameters due to the technical constraints of online dating platforms, while unintentional misrepresentation may also occur due to limits of self-knowledge \cite{ellison_managing_2006}. Generally, online daters share common expectations that everyone may lie or misrepresent themselves within online contexts \cite{ellison_profile_2012}. To enhance their attractiveness, individuals consistently modify their profiles based on their performance (i.e., whether they have successfully attracted those they are searching for), viewing the profile creation process as a constantly updating dynamic process \cite{whitty_revealing_2008}.

Furthermore, the degree of self-presentation exhibits variations when individuals present to different audiences based on familiarity and gender \cite{gosnell2011self}. Studies reveal that self-presentation motives are notably lower among highly familiar individuals of the same sex than less familiar individuals or those of the opposite sex \cite{leary1994self}. Additionally, self-presentation tends to be more modest when interacting with friends than strangers \cite{tice1995modesty}. Interestingly, when individuals anticipate meeting a potential dating partner for the first time, they often adjust their self-presentational behavior to align with the values deemed desirable by the prospective date \cite{rowatt1998deception,guadagno_dating_2012}.

\subsection{Profile Creation to Form Relationships}

Constructing effective profiles holds significant importance in forming relationships online \cite{lampe_familiar_2007}, especially in online dating platforms where less attractive ones are frequently quickly neglected due to the abundance of profiles to look through \cite{heino_relationshopping_2010,toma_separating_2008}. 

Previous research on online profiles has explored the role of profiles in forming online relationships, as people often make decisions about whether to interact further through signals and common ground in profiles \cite{lampe_familiar_2007}. Women were also found to base their decisions on the perceived attractiveness and nice personality of the potential match \cite{roshchupkina_rules_2023}. Also, young people not only simply present themselves to others, but also explore their personal identity, social identity, and gender identity through possible or idealized self-exploration in profile construction \cite{manago_self-presentation_2008}.

More research explored different profile elements such as free-text component \cite{toma_what_2012,toma_reading_2010}, photos \cite{hancock_putting_2009,lo_contradictory_2013,degen_profiling_2023}, and other factors like humor \cite{cantos-delgado_i_2023}. For example, in a study exploring how computers and humans could detect linguistic-related deception in profiles \cite{toma_reading_2010,toma_what_2012}, researchers found that while computerized linguistic analyses could detect deception, human judges were unable to accurately detect the trustworthiness of daters based on textual descriptions alone. Researchers also explored deception in photos and found that the authenticity evaluation of photos was positively related to that of free-text components in profiles while deception was positively related to the physical attractiveness of potential dates \cite{lo_contradictory_2013}. Interestingly, users may intentionally present worse personal profiles to lower their own attractiveness and form a good impression on attractive daters. Recently, researchers used serial picture analysis to explore self-presentation in online dating profile photos \cite{degen_profiling_2023}. They described the photos regarding the perspective, angle, body parts, posture, background, etc., and found 8 types of self-presentation: selfie, informative, snapshot, sociability, professional, incognito, suspending the subject, and challenging the logic. Their findings suggest that online dating self-presentation requires a combination of highly controlled (selfie) or staged photos (professional), romantic echos (informative) or the "playful nature of love" (incognito), authenticity (snapshot) and affability (sociability). Furthermore, an interesting study explored gender and culture-influenced humor employment in online dating profiles in the context of the UK and Spain \cite{cantos-delgado_i_2023}. While Spanish speakers may view humor as a risky tactic that could backfire, UK users favor it as part of their culture of not taking oneself too seriously.

Although previous research has explored the role of profiles in relationship formation and different profile components, how online daters create their profiles step by step, especially select photos, is still largely understudied. Therefore, our RQ1 aimed to explore the adaptive process of people selecting images and creating dating profiles.

\subsection{The Ideal Self and Actual Self}

Dating app users often grapple with the challenge of accurate self-disclosure and try to balance self-presentation pressures with the desire for authenticity \cite{toma_separating_2008,ellison_managing_2006}. Thus, they must often mediate carefully between these two contrasting motivations \cite{peng_be_2020}. For one thing, daters are aware that most relationships are still constructed offline, constraining them from faking too much in online profiles lest they meet the potential partner in person \cite{whitty_revealing_2008}. Also, people usually find more authentic profiles, which have low levels of selective self-presentation and high warranting value, can enhance trust and social attraction and thus increase positive outcomes in online dating \cite{wotipka_idealized_2016}. For another, individuals hold the intention of seeking partners who can understand and appreciate their actual selves instead of ideal versions of them \cite{swann_authenticity_1994}. Therefore, online daters tend to present an attractive but still real profile using equivocal statements \cite{ellison_profile_2012}.

In this context, previous research has identified several selves that occurred in self-presentation. In 1987, Higgins \cite{higgins1987self} defined three aspects of self: the \textit{actual self}, the\textit{ ideal self}, and the \textit{ought self}. The \textit{actual self} possesses all the traits that a person currently has, the\textit{ ideal self} has all the desired traits that a person wishes to have, and the \textit{ought self} possesses all the traits that a person ought to have based on social discourse. When creating an \textit{ideal self} online to impress others or just to explore possible selves \cite{manago_self-presentation_2008}, online daters often choose traits from their \textit{past, present, and future self} \cite{ellison_profile_2012}. In the online dating environment, individuals are influenced by limited cues \cite{ellison_profile_2012}, social norms that encourage the presentation of desirable attributes, market pressures inherent to dating platforms \cite{toma_separating_2008}, and peer pressure that often dictates profile content \cite{filter_dating_2017}. Consequently, online daters frequently portray a less authentic \textit{ought self} to attract potential matches \cite{filter_dating_2017}, although authenticity is valued in society and can enhance trust and social attraction \cite{wotipka_idealized_2016}.

Analogous to a theatrical performance, Goffman's analogy \cite{goffman1959presentation} introduces the concepts of "front stage” and “backstage,” where individuals strategically navigate between playing desired roles in the presence of others and authentic self-expression \cite{schlenker1996impression}. In this theory, the use of self-presentation is compared to a dramatic performance, where every individual is a performer in the show. Individuals strive to cater to societal expectations on the “front stage,” while the “backstage” area is where they isolate themselves from the audience, relax, and expose their real selves. Previous research also indicated that the self-presentational process occurs mainly in the “backstage” mode, and will only switch to a more active “front stage” mode on some specific occasions like important events or when others' perception matters \cite{schlenker1996impression}. When these important events occur, people manage their impression through \textit{given} expressions (conventional communication like spoken words) and \textit{given-off} expressions (presumed unintentional cues, such as nonverbal signals) on the "front stage" \cite{goffman1959presentation}. 

While previous work has extensively explored different selves in self-presentation, little is known about whether and how online dating app users present themselves with these selves in mind. Thus, our RQ2 aimed to explore how people present themselves with dating profiles and the reasons behind them.

\subsection{Authenticity and Deception in Profiles}

Due to the need to balance authentic self-presentation and marketing pressures in online dating \cite{toma_separating_2008}, authenticity and deception in profiles have become a key area of research. Researchers have explored what affects authenticity in profiles and found it positively related to self-esteem \cite{ranzini_love_2017}. Interestingly but somehow counterintuitively, higher educated dating app users showed lower self-esteem and were more likely to deceive in their profiles \cite{ranzini_love_2017}. Also, heterosexual users were found to be more authentic than homosexual, bisexual, and other users \cite{ranzini_love_2017}. 

Linguistic components, as important parts of profiles, have been explored in terms of deception. Toma et al. \cite{toma_reading_2010} first explored how linguistic patterns can be used to detect deception in profiles using computerized text analysis, then did a comparison between computers and human judges and identified specific linguistic patterns that suggest deception \cite{toma_what_2012}. For example, liars tend to use fewer self-referential words (e.g., "I", "myself") and more negative emotion words because these can help them psychologically distance themselves from deceptive behavior \cite{toma_reading_2010}. Moreover, human judges were unable to accurately detect the trustworthiness of daters based on textual descriptions alone as computerized linguistic analyses did \cite{toma_what_2012}. Instead, they mainly relied on linguistic cues unrelated to profile deception.

Previous research also addresses the authenticity issue in profile photos. While many online daters rate their own photos as relatively accurate, studies have found that their photos are often perceived by others as inaccurate \cite{hancock_putting_2009}. Physical attributes play a significant role in profile evaluations \cite{whitty_revealing_2008}, yet highly physically attractive photos are paradoxically deemed less authentic than those with lower physical attractiveness \cite{lo_contradictory_2013}. Interestingly, online daters tend to engage in more positive deception when interacting with highly attractive individuals, possibly to enhance their appeal \cite{lo_contradictory_2013}. This raises the intriguing possibility that users might intentionally downplay their own attractiveness in their profiles to create a favorable impression on attractive daters \cite{lo_contradictory_2013}. Furthermore, the authenticity evaluation of photos is positively related to that of free-text components in profiles \cite{lo_contradictory_2013}.

Gender difference is somewhat significant in profiles. While both females and males lie to enhance their attraction to potential dates \cite{lance1998gender}, different genders may lie in different aspects. For example, males often overestimate their height, while females often underestimate their weight \cite{hancock_truth_2007}. Regarding photos, female photographs appear to be less accurate and authentic than male photos \cite{hancock_putting_2009,lo_contradictory_2013}. Prior research \cite{guadagno_dating_2012} explored whether expectations about meeting impact deception in self-presentation under four anticipated meeting conditions: face-to-face, email, no meeting, and a questionnaire control condition with no pretense of dating. In this study, male participants were found to exaggerate self-presentation when anticipating more prospective future interactions with potential dates.

Although the authenticity and deception in profile photos and text are well explored, many studies have only investigated the extent and reasons for deception. There is still a lack of what specific profile elements may influence people's perception of others' profiles. Therefore, our RQ3 set out to discover the key factors that matter to people when evaluating other's profiles.

\section{Method}\label{sec:Method}

Semi-structured interviews were conducted with 20 online dating app users utilizing Tencent Meeting, Feishu, WeChat, and in-person. The goal was to explore the adaptive processes individuals employ when crafting their dating profiles, what desirable presentation they would like to achieve in profiles, and their perception of others' profiles. Participants were recruited through direct messaging and social media posts on platforms such as WeChat and Xiaohongshu. Only online daters who had experience with at least one dating application (including popular sites in both China and abroad, see Table \ref{tab:table1} and brief descriptions of these applications in Appendix \ref{DatingApps}) were selected. 

During the study, institutional research protocols were approved and rigorously adhered to. Ethical integrity was ensured, with informed consent meticulously secured from all participants, adhering strictly to anonymization protocols as dictated by the institutional Internal Review Board (IRB), thereby guaranteeing participant privacy. Upon completion of the interview, participants were compensated for their time with a 50 CNY shopping card.

\subsection{Participants}

A cohort of 20 participants (11 females, 9 males, see Table \ref{tab:table1}), who are college-educated young adults from various districts in China, was recruited for this study, following a purposeful sampling strategy \cite{suri2011purposeful}. All participants were native Chinese speakers aged 22 to 33 (average age: 25.0) having experience with at least one dating app (including popular sites in both China and abroad, see Table \ref{tab:table1} and brief descriptions of these applications in Appendix \ref{DatingApps}). The group presented heterogeneity in sexual orientation, including both heterosexual (N=17) and homosexual (N=3) individuals. Participants have varied educational backgrounds, ranging from undergraduate students to doctorate degrees, in fields such as public health, management, psychology, design, geography, computer science, and engineering.

The limitation here is all participants are Chinese. Although seven of them had living and educational experience abroad, certain limitations still exist (see the detailed discussion of the limitation in Section \ref{limitation}).

\raggedbottom
\begin{table*}[htbp]
\centering
\caption {\label{tab:table1} Demographic Information of Participants}
\resizebox{\linewidth}{!}{
\begin{tabular} {lcccccc}
\toprule
\textbf{ID} & \textbf{Gender} & \textbf{Age} &  \textbf{Sexual orientation} &  \textbf{Education} & \textbf{Experience with dating apps} & \textbf{Major}   \\

\midrule
$P1$ & Female & 24 & Heterosexual & Master's degree & Bumble, Tantan & Public health \\
$P2$ & Female & 24 & Heterosexual& Master's degree & Bumble, Tantan, Tinder & Public health \\
$P3$ & Female  & 23 & Heterosexual & Bachelor's degree & Bumble, Soul, Summer, Tinder & Public health \\
$P4$ & Female & 24 & Heterosexual & Bachelor's degree & Qing Teng Zhi Lian, Tantan & Management \\
$P5$ & Female & 27 & Heterosexual & Master's degree  & Soul, Tinder & Communication \\
$P6$ & Male & 27 & Heterosexual & Bachelor's degree & Qing Teng Zhi Lian, Soul, Tantan & Management \\
$P7$ & Male & 33 & Heterosexual & Doctorate degree & Bumble, Tantan, Tinder & Psychology \\
$P8$ & Female & 23 & Heterosexual & Master's degree  & Hinge, Qing Teng Zhi Lian & Design \\
$P9$ & Female & 23 & Heterosexual & Master's degree  & Qing Teng Zhi Lian, Soul & Design \\
$P10$ & Female & 22 & Heterosexual & Master's degree  & Qing Teng Zhi Lian, Soul & Design \\
$P11$ & Female & 26 & Heterosexual & Bachelor's degree & Jimu, Soul, Tinder & Design \\
$P12$ & Male & 24 & Homosexual & Bachelor's degree & Blued, Fanka & Geography \\
$P13$ & Female & 23 & Heterosexual & Master's degree & Jimu, Soul, Tantan & Design \\
$P14$ & Female & 25 & Heterosexual & Master's degree & Bumble, Jimu, Soul, Tantan, Tinder & Design \\
$P15$ & Male & 22 & Heterosexual & Undergraduate student & Tinder & Design \\
$P16$ & Male & 25 & Homosexual & Master's degree & Blued, Fanka & Engineering \\
$P17$ & Male & 27 & Heterosexual & Bachelor's degree & Bumble, Tantan, Tinder & Computer science \\
$P18$ & Male & 29 & Homosexual & Bachelor's degree & Blued, Tinder & Management \\
$P19$ & Male & 24 & Heterosexual & Master's degree & Soul, Tantan & Education \\
$P20$ & Male & 24 & Heterosexual & Master's degree & Soul, Tantan & Engineering \\
 \bottomrule
 \end{tabular}


 }
\end{table*}

\subsection{Interview Procedure}

Each semi-structured interview lasted thirty minutes to one hour. The interviews were conducted via Tencent Meeting, Feishu, WeChat, and in person when physically available \cite{strauss1997grounded}. The interviews were audio-recorded in Mandarin Chinese, and then transcribed into English after removing any identifying information and personal details to ensure confidentiality.

During each interview session, researchers first explained the purpose of the study and obtained informed consent from participants. Then, demographic information (e.g., name, age, gender, sexual orientation, educational background, and experience with dating apps) was gathered. 

\begin{figure}[htbp]
 \vspace{-0.3cm}
  \centering
    \includegraphics[width=.9\linewidth]{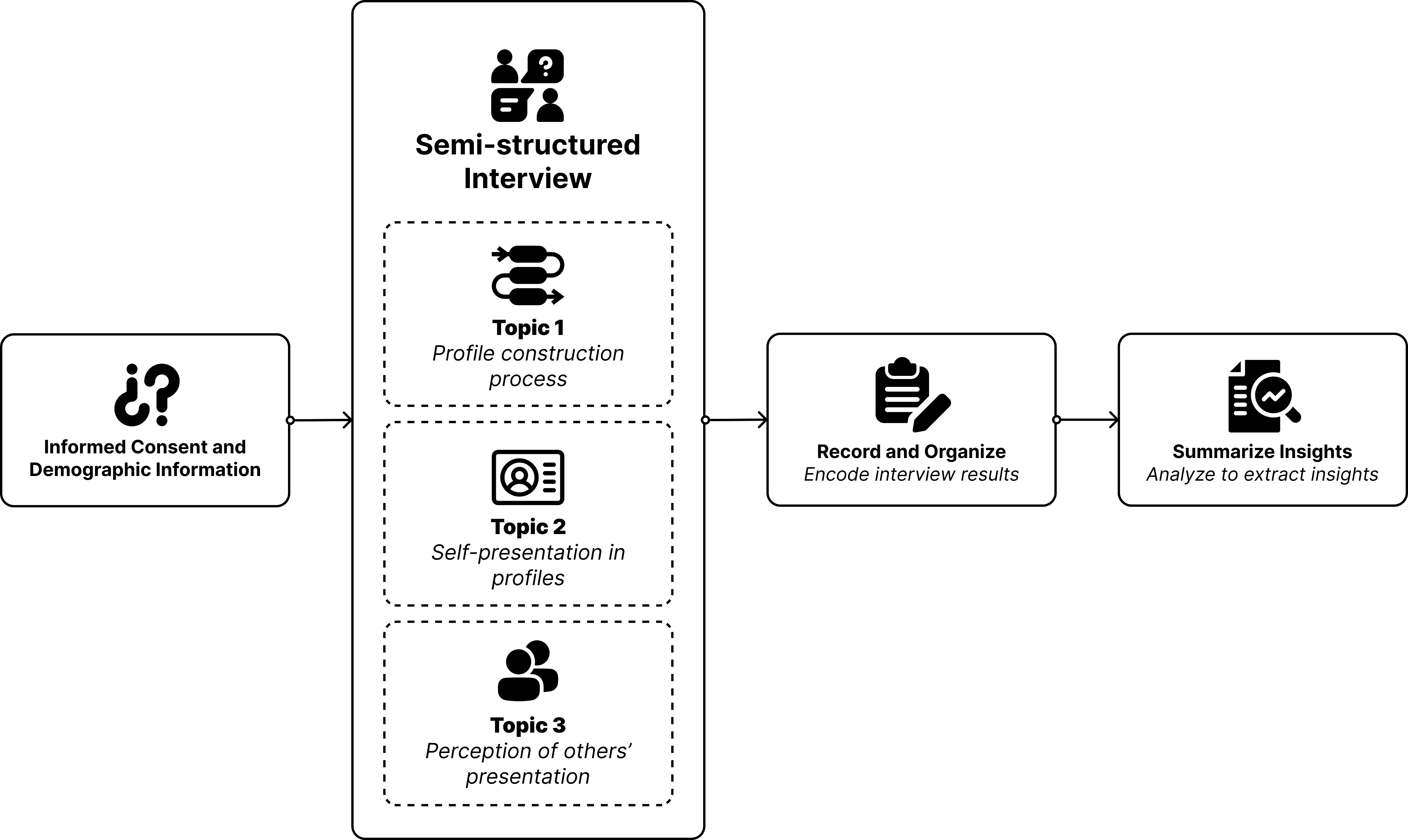}
  \vspace{-0.3cm}
  \caption{Interview process: We conducted semi-structured interviews with three main topics (profile construction process, self-presentation in profiles, and perception of others' presentations).}
 \label{interview_process}
  \vspace{-0.3cm}
\end{figure}

The semi-structured interviews were designed to cover three main topics (see Figure \ref{interview_process}): a step-by-step profile construction process to see how daters choose photos and text for their profiles (see the detailed process in Section \ref{ProfileConstruction}), questions regarding self-presentation in profiles, and perception of others' presentation. 

For the first topic, participants were asked to construct a new dating profile under the researchers' observation and provide their reasons for photo selection and text composition. Under each topic, specific questions were posed (see Appendix \ref{interview}), such as how participants select photos for their profiles, what information they choose to include or exclude, and their perceptions of physical attractiveness versus personal descriptions in achieving successful matches. These questions were intended to probe into their strategies and motivations in presenting themselves and assessing others on the platform.

\subsection{Constructing Profiles} \label{ProfileConstruction}

In this stage, we chose to use the profile-making interface of Tinder, one of the most popular online dating apps worldwide \cite{cantos-delgado_i_2023}, as a template. Prior to the interviews, researchers created a Tinder account for profile construction. Participants logged in to Tinder using the provided account. Then, they were asked to construct a new profile during a session with researchers, which was conducted either online via screen-sharing or offline in a secure environment. We took special care to protect participants’ privacy and data security. Crucially, we did not have participants upload their created profiles to Tinder or any other online platform as the final step. The session was audio recorded and only the final profile and selected photos were documented.

For online interviews, participants shared their screens with researchers, displaying the Tinder profile editing interface. This enabled researchers to view participants’ photo selections in real time as they navigated through their devices and uploaded images to their Tinder profiles. 

For offline interviews, the process was conducted in a private, secure setting where participants used a provided computer or their own device to create their Tinder profiles with one researcher. Researchers observed the participants’ actions as they selected images and uploaded them to their profiles. 

\begin{figure}[htbp]
  \centering
  \includegraphics[width=0.99\textwidth]{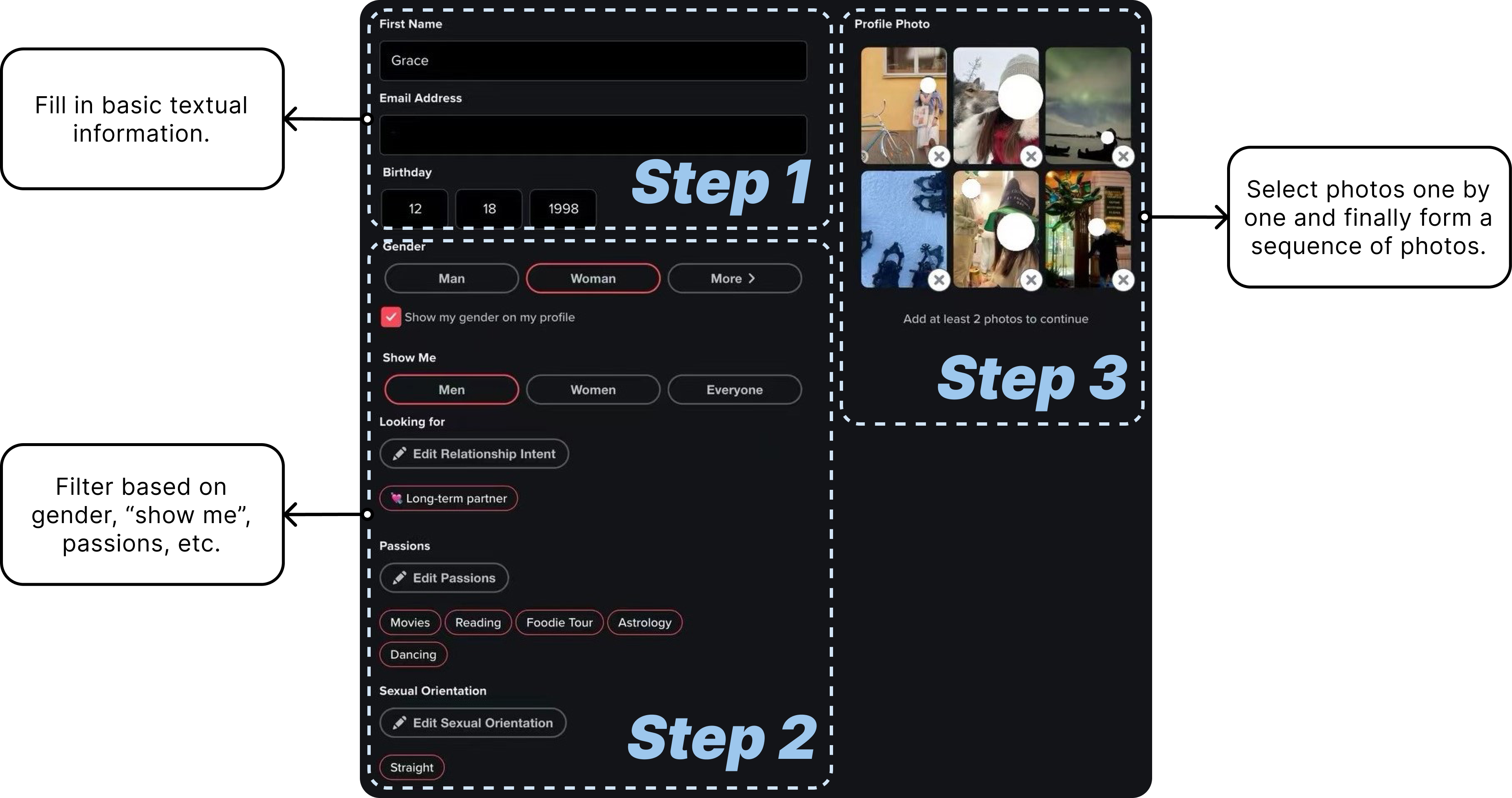}
  \caption{Profile construction process: We used Tinder as a template to guide users to create the profiles step by step. This is a profile example constructed using one researcher's photos following P4's photo selection and persona construction strategies.}
  \label{Profile}
  \Description{}
\end{figure}

Following the guide of the profile-making interface, participants filled in text information, filters (gender, relationship intent, etc.), and selected 6 photos step by step (see Figure \ref{Profile}). Participants were encouraged to think aloud when creating profiles and researchers further probed their reasons for text composition, photo selection (including types and sequence of photos), and choices for filters. 

Before participating, all individuals signed informed consent, fully aware that only selected photos and profile text would be documented and used in the research. Photos were anonymized, with no identifiable information linked to the images in our paper. We altered any distinguishing features in the descriptions to prevent identification. Additionally, participants were given the opportunity to review how their photos and text would be used in the final publication, and they could withdraw their consent at any time without any consequences. 

Upon the completion of the profile construction, we continued with the second and third topics of semi-structured interviews with the participants. However, as the study progressed, we noticed that participants primarily focused their discussions on the photo selection aspect. Consequently, we adapted our approach, shifting our emphasis to inquire specifically about their choices and reasoning behind the photos.


\subsection{Data Analysis}

Interview recordings were first transcribed and then open-coded by four researchers independently, following the guidelines of thematic analysis \cite{braun2006using}. This initial coding phase involved categorizing the data into preliminary codes. Researchers discussed these initial codes to check if they accurately reflected the data and refined and grouped similar codes into themes. Iterative discussion and coding continued for several rounds until researchers reached a consensus on the final themes. This process led to the categorization of the interview data into three themes: Visual Aspects of Photo Selection, Persona Construction Strategy, and Perception of Others' Profiles, which were presented in detail in Section \ref{sec:Results}.

\section{Results}\label{sec:Results}

In this research, we analyzed strategies for self-presentation and perception of others' profiles in online dating. While personal photos and display strategies were evident in the narratives of the participants, textual information was considered of less importance in this context. Therefore, we answered RQ1 in Section \ref{photo}, RQ2 in Section \ref{persona}, and RQ3 in Section \ref{perception}. Our findings not only clarified how users maintain authenticity while presenting themselves but also revealed how they adjust their presentation and communication strategies amidst changing social expectations and cultural differences. Through this analysis, we filled an important research gap in the existing literature about 
the adaptive process online dating users craft their profiles to attract potential dates, the way they present themselves with different selves in mind, and how they assess others' profiles through small signals in profile photos and text.

\begin{figure}[htbp]
 \vspace{-0.3cm}
  \centering
    \includegraphics[width=.99\linewidth]{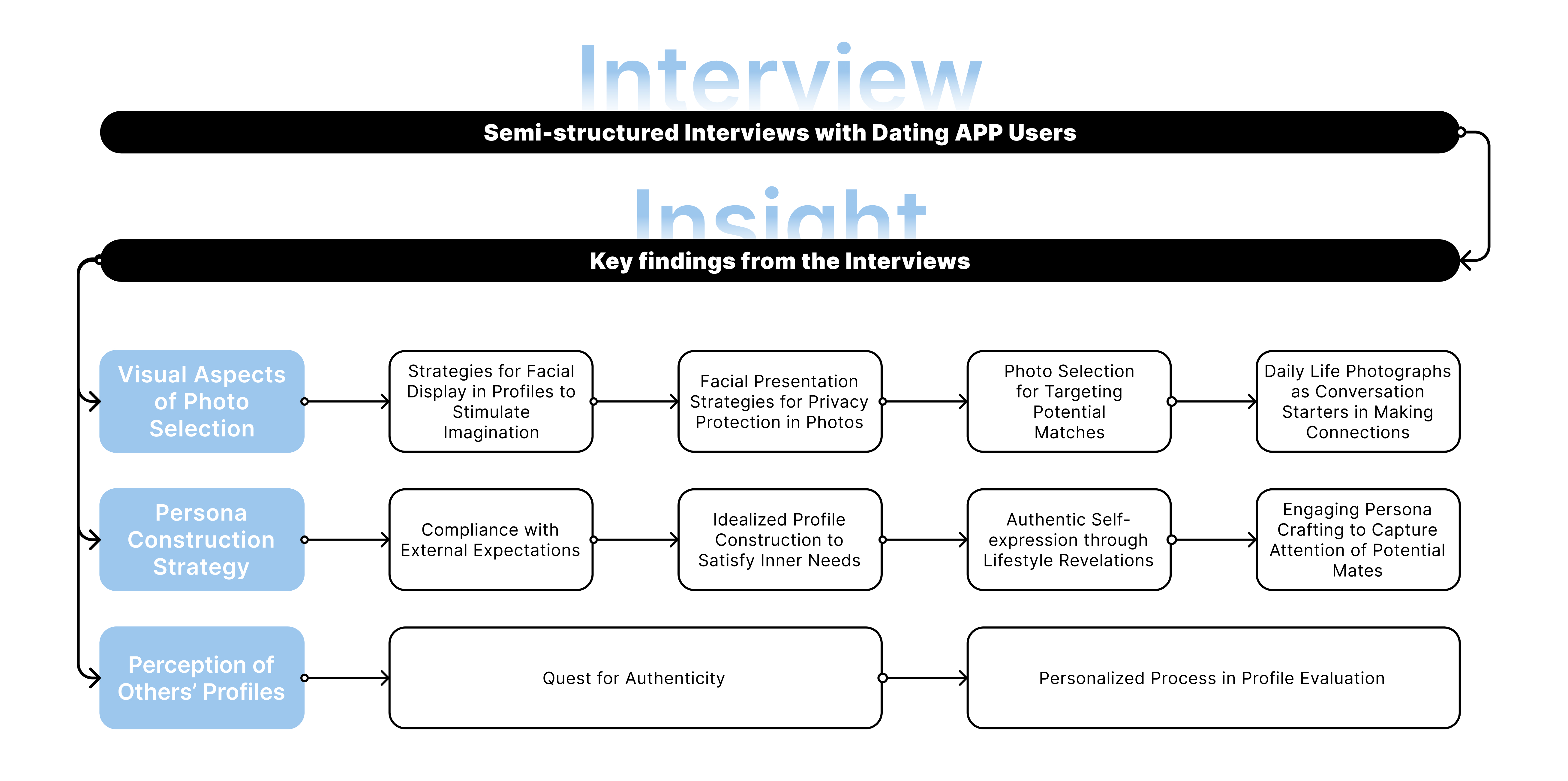}
  \vspace{-0.7cm}
  \caption{Key findings from the interviews}
 \label{key_finding}
  \vspace{-0.3cm}
\end{figure}

\subsection{Visual Aspects of Photo Selection \label{photo}}

\subsubsection{Strategies for Facial Display in Profiles to Stimulate Imagination}

Photo presentation plays a crucial role in online dating. Although the emphasis in online dating photos is usually on appearance and physique, participants in this study focused more on the atmosphere and vibe of their photos, sometimes even limiting the facial display. As participants P8 and P4 emphasized, their photo presentations were not just limited to simple facial displays. Instead, they focused on creating an artistic effect full of imaginative space through photography, aiming not only to showcase personal appearance but also to create an engaging and thought-provoking atmosphere.

\begin{quote}
    \small\itshape
    “I rarely posted highly edited selfies with just the upper body, often with uninteresting backgrounds like car interiors. Such photos lacked depth and failed to convey my interests or aesthetic taste. Instead, I focused on the atmosphere and storytelling in my photos, aiming to create a sense of interaction with the viewer.” (P8)
    \end{quote}

\begin{quote}
    \small\itshape
    “When I saw a photo on a dating app with a good layout and composition, I felt the urge to go in and zoom in. That photo really had a genuine feel. It had a certain vibe.” (P4)
    \end{quote}

Moreover, this strategy also includes considerations from P3, who intentionally avoided directly depicting their face or body to attract curious and genuinely interested potential dating partners.

\begin{quote}
    \small\itshape
    “I didn't post clear front-face photos of myself. Sometimes I posted a good-looking segment, but I wouldn't put it at the beginning; I chose to place it at the end. Because you had to swipe to see the end. If someone weren't interested in you, they wouldn't immediately see your face.” (P3)
    \end{quote}

\begin{figure}[htbp]
 \vspace{-0.3cm}
  \centering
    \includegraphics[width=.99\linewidth]{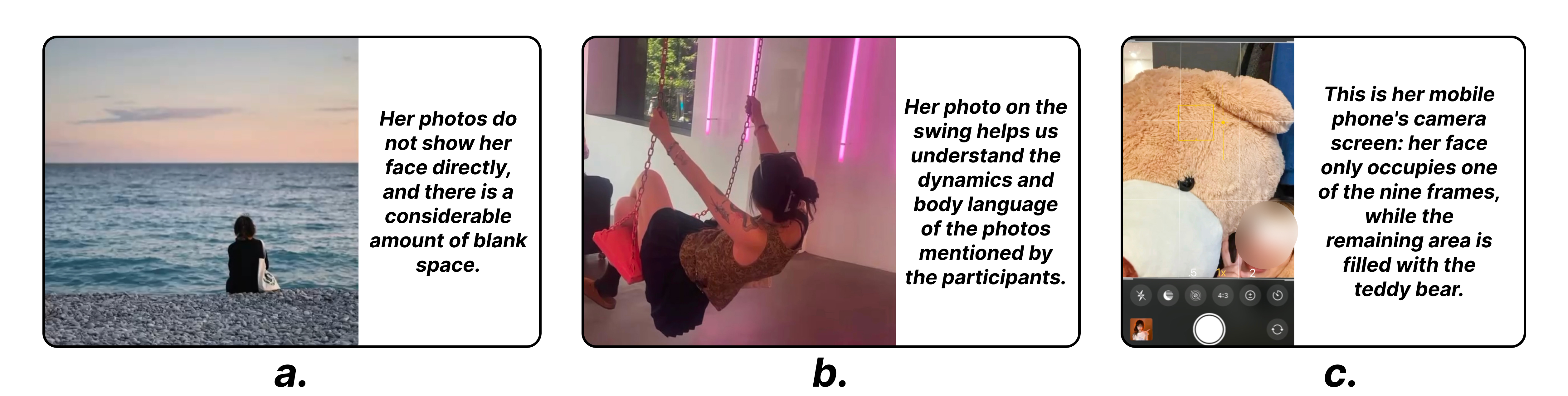}
  \vspace{-0.3cm}
  \caption{These are a few photos uploaded by P8 on the dating app.}
 \label{results1}
  \vspace{-0.3cm}
\end{figure}

Unlike some users who directly showcased their faces and bodies, this category of users adeptly employed elements like body language, lighting effects, negative space, and blurring techniques to add mystery and allure to their photos (see Figure \ref{results1}). This approach not only avoided direct judgment based on appearance but also stimulated the viewer's imagination and interest, thereby arousing deeper curiosity and attraction among potential partners. Through these strategies, participants successfully transformed their photos into effective tools for self-expression and communication, carving out a unique niche in the world of online dating.

\subsubsection{Facial Presentation Strategies for Privacy Protection in Photos}

Participants who preferred not to show their full faces in photos have developed strategies to protect their privacy while still giving potential dating partners a general idea of their appearance. P8 offered a cautious approach to facial exposure on social media, suggesting that people should limit facial exposure to a small part of the photo, preferably at the edge (see Figure \ref{results1}). P3 shared a similar viewpoint.
 
\begin{quote}
    \small\itshape
    “Really, when I took photos, the phone screen had nine grids, and my face could only occupy one grid. That is all the facial exposure I showed in the picture.” (P8)
    \end{quote}

\begin{quote}
    \small\itshape
    “I won’t show my full face, I chose to show half of it. It’s mainly for privacy concerns. I wouldn’t use a photo like an ID picture, but rather a side view.” (P3)
    \end{quote}

\begin{quote}
    \small\itshape
    “I wouldn’t put a photo like a graduation picture on such platforms. It’s important, but I’d rather share it with friends or family, not strangers. I don’t want all strangers to see me.” (P4)
    \end{quote}

Other participants expressed their reluctance to post full-face photos, fearing recognition by friends or family members who may also use the same online dating apps, which they found quite embarrassing.

Overall, these strategies represented a shift from traditional self-presentation methods on online platforms to a greater emphasis on privacy protection. By intentionally leaving space and using blurred or side-view photos, users reduced direct identification while still displaying their personal style and interests. This approach not only maintained individual privacy and avoided overexposure in public digital spaces but also effectively prevented awkward encounters with acquaintances on dating platforms. This focus on privacy and careful self-presentation was key to establishing healthy and secure communication in the digital dating environment.

\subsubsection{Photo Selection for Targeting Potential Matches}

Photo selection on online dating platforms was strategic for screening and attracting potential dates. As P9 articulated, the first photo in social media profiles played a crucial role, requiring artistic and design appeal to attract viewers' attention. Serving as the "facade" of personal profiles, the first photo not only reflected an individual's taste and interests but also served as a starting point for potential communication and interaction. Thus, it became essential to choose an impactful and intriguing first photo.

\begin{quote}
    \small\itshape
    “When I used dating apps, you generally saw one person at a time, and they only showed the first photo initially. If interested, you swiped to see the rest of their photos. So, the first impression was really important.” (P9)
     \end{quote}

Additionally, as P13 pointed out, individuals tended to select visually appealing photos for their profiles, often leading to heavily edited or exaggerated images through photo editing software. While these edited photos might have immediate enhanced visual appeal, they could also lead to unrealistic expectations and misrepresentation, potentially affecting the authenticity of online interactions and relationships. Notably, P13 and P16 prioritized the aesthetic appeal of photos over the accuracy of personal representation.

\begin{quote}
    \small\itshape
    “You really couldn't trust all the beautiful photos on dating apps. Many people were deceiving, and they didn't look like their photos in reality. I did the same; I didn't care if my friends said it didn't look like me. Even if they sometimes said I'd edited it too much, I still posted it if I thought it looked good. People on online dating didn’t know what you looked like anyway.” (P16)
    \end{quote}

Some participants preferred a contrasting strategy in photo arrangement. As P2 and P3 mentioned, if the first personal photo was particularly stunning, subsequent photos might include humorous or unconventional images to balance the distance created by physical attractiveness and create a contrast effect.

\begin{quote}
    \small\itshape
    “Finally, I thought it was important to create a contrast. Although I might have looked good, it could create a sense of distance. To reduce that, I might have added some humorous or quirky photos to balance that distance.” (P12)
    \end{quote}

P14 also mentioned adopting a strategy to attract people interested in specific types of photos, such as selecting sports-related images to attract those interested in sports, thereby initiating conversations based on common interests.

Overall, these findings highlighted the complexity and strategic considerations involved in selecting photos on dating apps. Users carefully crafted their profiles not only to present themselves in a visually appealing way but also to attract specific types of matches, initiate conversations, and manage perceptions of their personalities and interests.

\subsubsection{Daily Life Photographs as Conversation Starters in Making Connections}

Daily life photos played a significant role in maintaining conversation and interaction in online dating. As recommended by P9, photos in a personal profile should emphasize an individual's daily life and interests, such as food and entertainment activities.

Photos of everyday life provided a relaxed and authentic starting point for communication, making it easier for other users to engage and foster connections based on shared interests and daily experiences. P2, P3, P5, and P15 mentioned that compared to sharing movies, books, or text, sharing relatable everyday experiences like food would be more suitable as a conversation starter. Such shared experiences could help find potential cultural links, such as common lifestyle habits or preferences.

\begin{quote}
    \small\itshape
    “I usually avoided deep or controversial topics in initial conversations. I preferred discussing light-hearted subjects like food and drink. For instance, I might have shared photos of myself with my favorite food, sparking conversations about the best places for that food. This approach avoided serious debates and easily found common ground. If the other person hadn't tried the food, they could taste it and continue the conversation with me, naturally leading to more similar topics.” (P9)
    \end{quote}

Moreover, the perspective of P16 added another dimension, indicating that some users could be reluctant to share their thoughts and mental states with strangers in the initial stages. Such deeper sharing usually occurs after some progress has been made in the relationship. 

\begin{quote}
    \small\itshape
    "I tended to share... for example, I said I liked to sing and that was all I would share. It just didn't go on anymore about what traits I had in that area, so if you wanted to know about it, you could just chat about it later." (P16)
    \end{quote}

In summary, the use of everyday life photos on online dating platforms not only reflected an individual's lifestyle and interests but also provided an effective way for users to establish connections based on shared experiences. This strategy not only presented one's preferences but also laid the foundation for more profound future interactions.

\subsection{Persona Construction Strategy \label{persona}}

Our participants utilized various strategies to construct different personas in order to align with external expectations, respond to internal desires for traits that they wish to possess, express themselves authentically, or capture the attention of potential mates.

\subsubsection{Compliance with External Expectations.}

According to participants, they actively molded their online personas to align with external societal and potential partners' expectations and requirements. 

Participants noticed the cultural differences between countries, tailoring their profiles to the requirements of different social environments. P1, who had online dating experience both in China and Sweden, mentioned a significantly different self-presentation strategy meeting the requirements of different social environments. When she was in China, she deliberately avoided mentioning her size because she was considered a bit overweight. But she was not afraid to list it in Sweden because her weight was considered average there. 

\begin{quote}
    \small\itshape
   "I think it may be because the standards in China and Sweden are different, because they may not care much about body shape (in Sweden)...girls who are considered very plump in China look just perfect here, so I don’t care much about this aspect. But in China, I would be more concerned about it." (P1)
\end{quote}

Meanwhile, although self-identified as introverted, P1 emphasized her outgoing side by exaggerating her passion for nature and outdoor activities when she tried to use dating apps in Sweden. She discovered that most people showed this trait or related information, which made her believe that people there would expect to find cheerful and outgoing partners. 

\begin{quote}
    \small\itshape
   "I think because each society values different points, I will make some appropriate adjustments and beautification based on this." (P1)
\end{quote}

P4, who intended for a serious relationship, would choose more conservative photos with less skin exposure when she was in China. She explained that people would expect and like those dressed more conservatively; otherwise, they would be considered looking for only sexual relationships.

Participants highlighted the need to consider potential partners’ expectations with different relationship intents. Those who intended for serious and long-term relationships, like P4, often carefully choose photos with less skin exposure and "sexual" signals. 

\begin{quote}
    \small\itshape
   "If you use photos revealing more skin on dating apps, people will ask you if you want to go for sex." (P4)
\end{quote}

P4 added that photos revealing more skin would potentially attract those looking for “short-term fun” in such a conservative society. With the intention of looking for a relationship that could be "the source of security and power of support," P12 would consider what kind of profile might attract more people when crafting his profile. He also commented that photos containing “sexual” signals might raise the “swipe right” rate, yet they would attract more superficial relationships. Similarly, P14, who was looking for a long-term relationship, would carefully construct a high-quality profile, choosing photos representing her characteristics instead of those with more skin exposure. On the contrary, P15 highlighted the necessity of showing attractiveness or openness when looking for a “sexual” relationship.

\subsubsection{Idealized Profile Construction to Satisfy Inner Needs.}

Apart from adjustment according to external requirements, participants were aware of the traits they wished to possess or were looking for, crafting a profile that mirrors their "ideal" self and the traits they sought in a potential partner. 

Participants tended to exaggerate the parts of themselves that they would like to idealize. P12 said, “If you want something in your heart, you may exaggerate it.” He noticed the unconscious misrepresentation in his profile construction process and believed it had to do with “self-perception bias.”
\begin{quote}
    \small\itshape
   "People tend to think better of themselves. I feel like in terms of weight, this may be a point that I am not very satisfied with. I may not be able to change it in reality. Maybe I have always imagined a lighter me in my mind. I might be inclined to write this weight down a bit (in profile)." (P12)
\end{quote}

For example, P12 was unsatisfied with his weight. He appeared to be aware of constructing an "ideal self" with a lower weight. He argued, “It's not intentional. It may be subconscious. It's a habit. I may have lied to myself for too long and have taken it as reality.” When he crafted his online profile, he also tended to present himself with a lower weight. He added that he was still real because it was an “unfalsifiable proposition and deviation within a reasonable range” and thus couldn’t be found out by others.

Participants would also design their profiles according to the ideal partners they were looking for. As P1 said, she wanted to find someone with similar hobbies and lifestyle. Her profile presented her habits (e.g., drinking, smoking) and hobbies (e.g., cooking, photography). 

\begin{quote}
    \small\itshape
   "I like people who are smart, and I also hope that I can show that I am very smart." (P1)
\end{quote}

She specifically highlighted her education and stated she had the same criteria for self-presentation and looking for ideal partners. Therefore, she would show the same traits with the same level she desired in potential partners in her own profile in order to find "a perfect match". Similarly, P3 and P4 mentioned that they would present their hobbies and interests to find those that resonate with them. Furthermore, P12 went to a university ranked 73 in China, but to highlight his academic prowess, he said he went to one of China's top 10 universities on his profile. He explained that his “cult of academic qualifications” made him write down a better university to attract people with better education, though this time, he feared being exposed.

\subsubsection{Authentic Self-expression through Lifestyle Revelations.}

Although misrepresentation or exaggeration existed due to external or internal needs, participants still wanted to express themselves authentically in profile construction, using online dating platforms as stages for self-expression and expressing the interesting moments or happenings in their lives.

Participants emphasized their realness in profile construction. P13 mentioned that she would carefully keep authenticity even when choosing photos purposefully for persona construction. The dating process was deemed an “art” by P12, while designing profiles considering the preferences of his “ideal partner” was too mechanical and formulaic. 
\begin{quote}
    \small\itshape
    "I think it (designing profiles) is too complicated...I have to assume the type of person I like and what kind of person he might like. Then I need to match my presentation with his preferences. I think it's a little too by the book." (P12)
\end{quote}
Thus, he created his profile in a more casual and real way instead of making complex assumptions, putting whatever he wanted to show. He also reported that his profile was quite consistent with himself in real life. P15 agreed that considering others’ opinions was too much pressure and described his profile as “very objective without any subjective description.”

\begin{figure}[htbp]
 \vspace{-0.3cm}
  \centering
    \includegraphics[width=.99\linewidth]{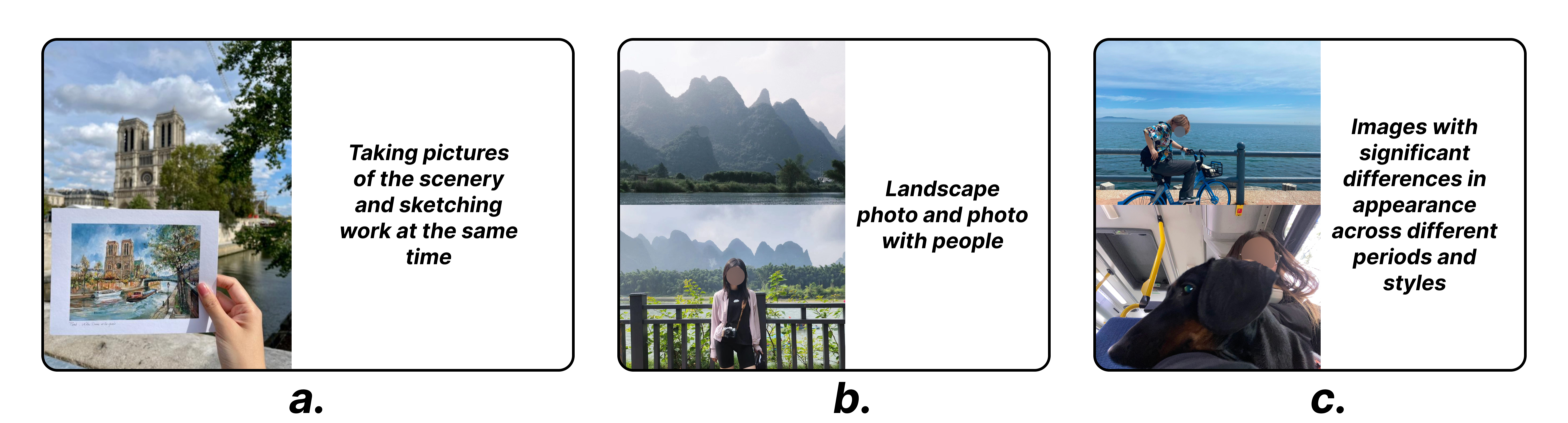}
  \vspace{-0.3cm}
  \caption{Example photos according to participants' statements: \textbf{a}. Use "pictures of the scenery and own sketching work" to indirectly show one's talent. (P8) \textbf{b}. Photos of landscape that one shot could express a moment moved one person, while photos of oneself simply showed one's appearance. (P4) \textbf{c}. P3 provided two photos with varied appearances and hairstyles taken from different periods of time, which aroused others' curiosity.}
 \label{results2}
  \vspace{-0.3cm}
\end{figure}

By revealing their lifestyle and talents, participants expressed their authentic selves and interesting souls. P14 would display her life in a more real and natural way rather than meeting others' expectations or being too ostentatious. She incorporated fitness into everyday photos to present her characteristics and lifestyle in a more synthetic way instead of focusing on her shape only. P8 held a similar opinion that presenting interests and talents on online dating platforms should not be a simple skill demonstration, such as "taking selfies directly in the mirror of the gym." Instead, it should incorporate skills or hobbies into real life, such as "wearing fitness clothes in everyday photos". She took "taking pictures of the scenery and your own sketching work at the same time" as an example, highlighting the interesting aspects brought by indirect lifestyle display (see Figure \ref{results2} a).

\begin{quote}
    \small\itshape
    "In short, you cannot just show that you are learning a certain interest, it should be extended to the fact that your interests have affected and been integrated into your life. The starting point is that you love life. This is the ultimate goal of talent presentation on dating platforms." (P8)
\end{quote}
In this way, interests and talents could be displayed more naturally and attractively as important parts of an individual’s life. Additionally, P4 highlighted the difference between two types of photos during a trip: photos of oneself and photos one took (see Figure \ref{results2} b).

\begin{quote}
    \small\itshape
    "Because when you shoot a landscape, it means that you sincerely think that moment is beautiful, and it touches you. I think this is different from posing for a picture of a person." (P4)
\end{quote}

By showing photos one took during a trip, P4 wanted to express a moment that touched her. She said, “I am not telling people that I love the world in a pretentious way, but I am really expressing my love for some creatures and things in the world.”

\subsubsection{Engaging Persona Crafting to Capture Attention of Potential Mates.}

Additionally, some participants utilized special strategies to capture the attention of potential mates, such as showing niche and arcane hobbies and using contrasting photos.

By presenting niche hobbies and interesting text, P4 aimed to craft a mysterious persona to attract potential mates. For example, she would mention astrology as one of her hobbies in profile construction. 

\begin{quote}
    \small\itshape
    "You can choose one or two unpopular ones there because you might actually meet someone with similar hobbies. It (astrology) also occupies a little bit of a niche. I think it will be a bit mysterious. Maybe people think it is very good and they will ask you." (P4)
\end{quote}

She believed that presenting unpopular hobbies would not only catch others’ eyes but also help find those who possess similar hobbies. Besides, she mentioned that she would write interesting text, like “only talk until 10 pm,” to also craft a niche persona. P4 believed that most people would only use dating apps late at night. Thus, a person who slept early would catch other’s attention immediately.

\begin{quote}
    \small\itshape
     "I will only use dating apps after I have finished my work during the day. Otherwise, I would not just open it and use it casually. I think most people should be like this. We use “only talk until 10 pm” to create a persona that goes to bed early, which is a very niche persona. Don't you ever wonder why a person goes to bed so early?" (P4)
\end{quote}

Similarly, some participants would strategically present their photos to make people curious. For example, P10 would choose profile shots or back-view photos to maintain a sense of mystery and be more imaginative. P18 and P19 had a similar strategy that avoided exposing their face in photos. P3 mentioned that using contrasting photos might arouse curiosity, taking her photos with varied styles from various periods as an example (see Figure \ref{results2} c). 

\begin{quote}
    \small\itshape
    "I can find you a photo that was taken a long time ago...a Swedish guy said that when he first saw that photo, he thought that it was a fake photo and I used fake photos to deceive people...because at that time I had short and orange-red hair...my hair is no longer orange...then they asked if it was me." (P3)
\end{quote}
P3 indicated that she didn't deliberately put contrasting photos to capture others' attention, yet it turned out that it worked quite well.

\subsection{Perception of Others' Profiles \label{perception}}

\subsubsection{Quest for Authenticity}

In our study, a key aspect of participant behavior was their cautious approach toward assessing the authenticity of others' profiles. This caution was particularly evident in their reactions to profiles, where the specific properties in a photo, including displays of material wealth or the quality of the pictures, significantly influenced their perceptions.

Participants like P15 and P1 prioritized authenticity, favoring realistic and relatable images over those appearing overly perfect or staged. For P1, overly perfect photos conveyed a message that the person had some purposes other than looking for a relationship, such as attracting followers to be a celebrity. Rather, she would like to see profiles with everyday photos and sharing of one's life. 

\begin{quote}
    \small\itshape
    "If his photos are all very beautifully edited, and each of them looks like they were cut out of a magazine, which makes me feel as if he didn’t use this app to match others...some people will even post their social media accounts below, and I will think that maybe he wants to be a celebrity." (P1)
\end{quote}
Many participants demonstrated a careful approach when encountering profiles with overly perfected images or those that exhibit significant inconsistencies among various profile pictures. P1, for instance, took a holistic view of the profile, evaluating its text in addition to focusing on just the appearance to validate the authenticity of the profile.   
\begin{quote}
    \small\itshape
    "(I) will assess text besides photos. It also depends on what content...an introduction to yourself, which may be more comprehensive and detailed. Be able to judge this person’s hobbies, interests, and what he values. These make it more real to me." (P1)
\end{quote}
P13 exemplified this act of caution, especially towards profiles with photos that seemed excessively perfect or of high quality. To ascertain their authenticity, P13 often saved these images and utilized tools like Baidu Image Search or consulted friends. 
\begin{quote}
    \small\itshape
    "I will directly search on Baidu, save the image, and then use Baidu's image recognition. Alternatively, I ask friends, 'Do you think this person looks like someone from a photo online?'" (P13)
\end{quote}

P12 also highlighted a common skepticism towards overly polished or inconsistent photos. He would often evaluate the consistency of all photos to check whether this was a real person or not. He said,  "If the main picture and the subsequent ones in their profile are nearly identical, it is probably their real image. But if there’s a stark difference between the first and the following pictures, they are likely images from the internet."

Reflecting on the interviews, a pattern emerged in how participants discern deception from profiles based on specific properties in photos. P5, P11, and P13 consistently expressed aversion to profiles featuring luxury brands, vehicles, and other symbols of affluence, which they viewed as disingenuous and showy. 

\begin{quote}
    \small\itshape
    "What I'm not interested in are... pictures with more than five brand logos, blurred ones, like a photo taken inside a Lamborghini, wearing Dior and a Rolex watch." (P5)
\end{quote}


P14 felt unreal, especially when she saw low-quality or obviously copied images. P1 had a similar feeling when she saw photos that were "too blurry", which made her feel fake. These insights suggested that for these participants, authenticity was tied to realism and consistency in online presentations, while blatant displays of wealth or poor-quality images raised suspicions about a profile's genuineness. 

Interestingly, P4 mentioned a specific aspect when she felt a person was unreal, "It feels like the facial features are patchworked...no, not patchwork, maybe because there is no light and shadow transformation." She felt that the light and shadow effect could be affected significantly by photo editing, which made it rather important in assessing the authenticity of photos.

\subsubsection{Personalized Process in Profile Evaluation}

When evaluating profiles, our participants had different processes and strategies. While physical attributes were the predominant focus of users, text descriptions in profiles were deemed by some participants to better reflect characteristics.

Most participants (except P9, P11, P13, P17, P18, P20), both female and male, mentioned that their first impressions were built on photos in which the most important factors were considered physical attributes (especially appearance and height). P1's remarks underscored the significance of appearance in online dating.

\begin{quote}
    \small\itshape
    "The first thing I do when opening a profile is to look at the person's photos, especially the profile picture, as it's the first thing I see. The age and height information, otherwise usually displayed in the bottom left corner of the photo, are also crucial factors in my assessment." (P1)
\end{quote}

P14 echoed a similar sentiment: "I first look at the face, and if it seems okay, I then check the height. Only if both these aspects are satisfactory do I proceed to view other photos and read the personal introduction." P14's approach revealed how users in online dating sequentially filter potential matches, starting with physical characteristics. P2 and P4 also emphasized the importance of facial attractiveness. P2 candidly expressed, "If the face looks good, I continue browsing; if not, I stop." P4 shared a similar view, "For me, the main focus is on the face. I don't spend time reading their lengthy text descriptions." P19 further validated this perspective, "My primary focus is on the person's face and body. Their photos are key to my assessment."

Apart from physical attributes, some participants mentioned that photos containing animals or unique content would also attract them. P1 underscored the importance of showing animals in profile. She said, "I think photos with small animals will greatly increase my probability of matching with this person by 80\%. I personally love animals very much. I think if he particularly likes animals, he must be a good person." P3 shared her experience when evaluating a profile with unique content:

\begin{quote}
    \small\itshape
    "For example, I saw a boy who was in Egypt. He was Swedish, but he was working in a relatively poor country, also (doing) that kind of (animal) protection. Then, the first photo on his homepage was a baby turtle on his face. He was on the beach, and then I felt I had to talk to him to find out what he had experienced. We talked about his work on sea turtle protection." (P3)
\end{quote}

Some participants, like P3 and P4, expressed their preference for photos over text. For P3, photos could convey more information (e.g., environment, person, and vibe) and be more reliable due to the higher threshold and cost for photo creation. This perspective was confirmed by P4, "You can plagiarize. You can just quote some good sentences or something. I think it loses some authenticity."

\begin{quote}
    \small\itshape
    "Because the creation of photographs is of a higher threshold than the creation of words. If you really want to show you are into puppies, then you must have a picture of a puppy. But if you don't have a picture of a puppy, you just write a text saying I like puppies...photos are more believable than words." (P3)
\end{quote}

Meanwhile, participants mentioned that text could reflect one's real characteristics and personalities better and more easily than photos. P3 said, "Because words can be more subtle than photos...Words will easily reveal that you are a misanthrope or that you are carpe diem people." 

Interestingly, while most participants valued photos more than text, some of them would specifically look for attractive cues in profile descriptions after photo evaluation. P2 described one special text description that caught her attention, "One day, I saw a boy’s profile saying that he could easily be moved and that anything could make him cry...I think he described it very well." For P15, interesting or unique text would attract him. He gave an example like "30 years old, divorced, having two kids", which was Internet slang in China that was popular a long time ago. 

\section{Discussion}\label{sec:Discussion}

Our study investigated how experienced online dating platform users strategically craft their profiles. The findings of our study shed light on the intricate process of online dating profile creation, focusing on visual aspects of photo selection (RQ1), persona construction strategies (RQ2), and perception of others’ profiles (RQ3).

\subsection{Key Findings}

\subsubsection{Establishing Common Ground through Signals in Photo Selection.}

Our findings revealed that online daters carefully select their profile photos to stimulate imagination, protect privacy, target potential matches, and make common connections. 

We relate our findings to previous research on how Facebook profile elements influence relationship formation in online communities \cite{lampe_familiar_2007}. Researchers described a theoretical framework incorporating signaling, common ground, and transaction cost theory to understand and determine which profile elements matter in forming online connections. As they noted, generally, only conventional signals that are easy to produce and manipulate are supported in online contexts. Therefore, signals can be more important in a reduced-cue online environment \cite{ellison_managing_2006}. Moreover, people tend to make connections with those who share common ground (locations, interests, and referents) with them, as the common ground can reduce the cost of connection \cite{lampe_familiar_2007}. 

In our study, how online daters selected photos aligns with this theoretical framework but with a slight difference. Although Facebook users often use common referents as common ground \cite{lampe_familiar_2007}, online daters often need to face people they do not know or have no social connections. In other words, Facebook connections are built upon the shared social network offline, while online dating relationships are formed in a larger "market". Without common referents, online daters tend to establish common ground through profile photos that are more likely to capture initial attention compared to text \cite{van2022people,scott2016motivation,seidman2013effects}. They carefully depicted their profiles through different signals in photos and tried to build common ground by using photos from their daily lives. As previous research \cite{lloyd_identity_nodate,fiore_homophily_2005} has pointed out, online dating users prefer people similar to themselves, especially in attributes related to daily life. For instance, P9 mentioned that she would display easily shared elements (e.g., her favorite restaurant) in her photo messages. If a potential date has been to this restaurant, this can be a shared experience; if the person has not been to the restaurant, it can still be relatively accessible for them to visit together. Thus, sharing aspects of everyday life can be an easily established cultural connection. While trying to target potential matches, our participants also mentioned the importance of common interests (e.g., sports) as an instance of building common ground in photo selection. This echoes previous findings saying that people use informative type of photos to show lifestyle and romantic ideals \cite{degen_profiling_2023}.

Similar to the role of common referents as verification in Facebook \cite{lampe_familiar_2007}, common connections can also assist in determining the accuracy of information provided by potential dates. Users often need to thoroughly search for similar cultural connections to assess the authenticity of a potential mate's profile. As P1 pointed out, examining a potential date's life interests and values can aid in constructing a more authentic portrayal of that person. This approach not only helps users find potential mates with shared interests but also enhances opportunities for mutual understanding and trust-building. Consequently, users can gain a deeper insight into the personality and values of their potential mates, enabling them to make more informed decisions.

\subsubsection{Incorporating Desired Traits into User's Own Profiles.}

One of the most interesting findings in our study is that online daters tend to construct their profiles mirroring the "ideal self" and the "ideal partner". For example, P12 was unsatisfied with his weight and always desired a lighter version, and he put that desired lighter weight in his profile. P1 desired a smart potential partner, and she also wrote that she went to a good university. 

Previous research has identified several selves in terms of self-presentation. Higgins \cite{higgins1987self} defined three aspects of self: the \textit{actual self} refers to who a person currently is, the\textit{ ideal self }refers to who a person wishes to be, and the \textit{ought self} refers to who a person ought to be based on the social discourse. Moreover, individuals also consider the \textit{past, present, and future self} in their self-presentation \cite{ellison_managing_2006,ellison_profile_2012}. For example, individuals often choose traits from the \textit{past, present, and future self} to represent themselves and create an \textit{ideal self} \cite{ellison_profile_2012}. People tend to present an \textit{ideal self} online to impress others or just to explore possible selves \cite{manago_self-presentation_2008}. In the online dating environment, daters are influenced by limited cues online \cite{ellison_profile_2012}, social norms that require them to present attributes others appreciate and value, market pressure inherent on dating platforms \cite{toma_separating_2008}, and peer pressure that often helps decide what to include in their profiles \cite{filter_dating_2017}. Therefore, although authenticity is valued in society and can enhance trust and social attraction \cite{wotipka_idealized_2016}, online daters often present themselves with a less authentic \textit{ought self} to get potential matches \cite{filter_dating_2017}.

\begin{figure}[htbp]
 \vspace{-0.3cm}
  \centering
    \includegraphics[width=.5\linewidth]{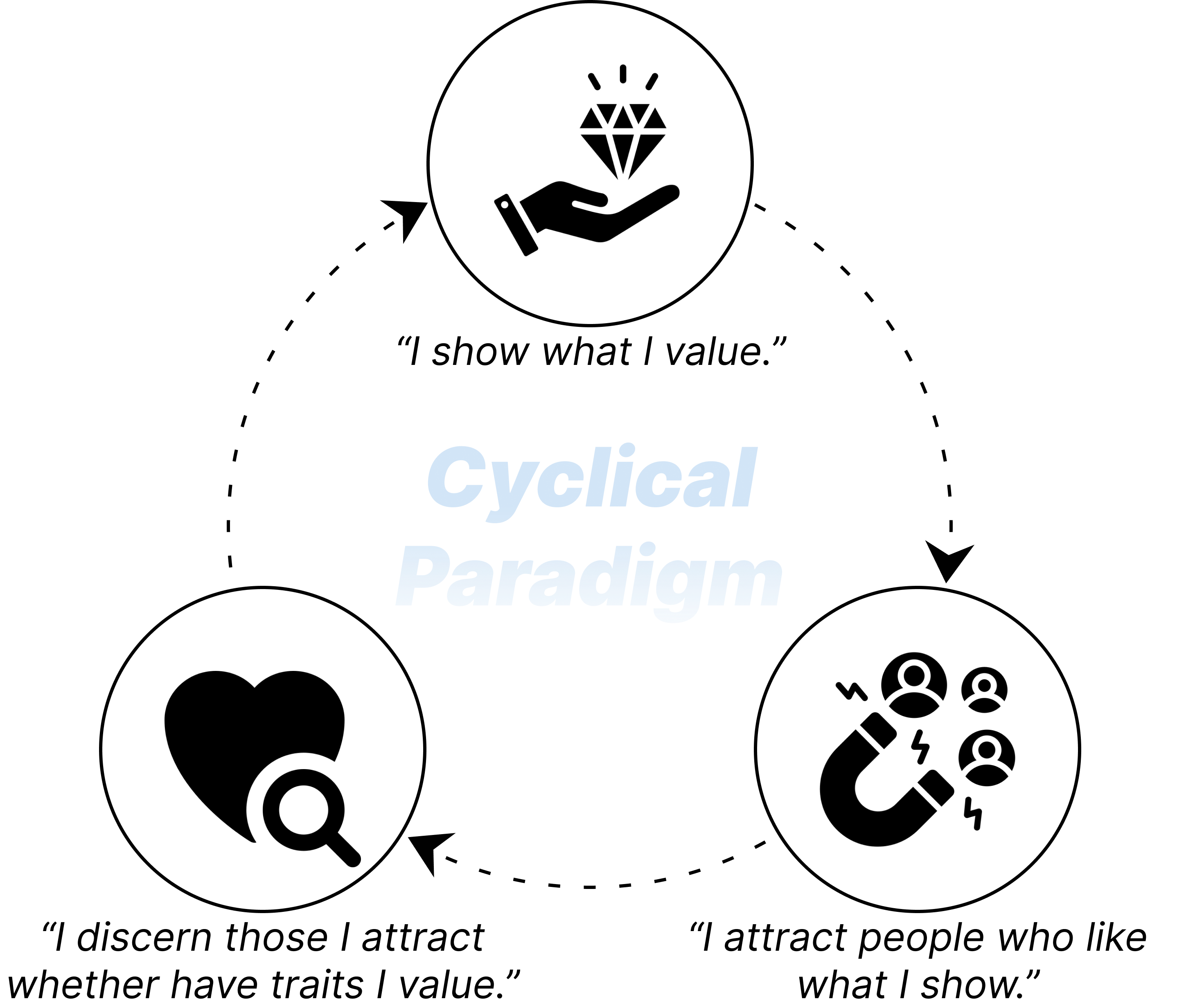}
  \vspace{-0.3cm}
  \caption{The cyclical paradigm when users create profiles: "I show what I value; I attract people who like what I show; I discern those I attract whether have traits I value."}
 \label{paradigm}
  \vspace{-0.3cm}
\end{figure}

In our study, participants responded to external societal expectations and potential partners' requirements, fostering the presentation of an \textit{ought self} \cite{higgins1987self,filter_dating_2017}. They also created an \textit{ideal self }by incorporating some traits they wish to possess \cite{ellison_managing_2006,manago_self-presentation_2008}. However, the reason for constructing profiles mirroring the "ideal partner" may be different from the one to present an ought self due to external forces \cite{filter_dating_2017}. Participants first identified their internal desires and expectations for the "ideal partner", and then showed the aspects they value on their partners in their own profiles. By doing this, they believed that they would attract people who like these attributes, and they would also check whether those attracted by them have these traits. This psychological process by which online dating users create profiles can be conceptualized as a cyclical paradigm (see Figure \ref{paradigm}). This paradigm can be expressed as: "I show what I value; I attract people who like what I show; I discern those I attract whether have traits I value." This cyclical process encapsulates the dynamic interplay between personal values, self-presentation, attraction patterns, and subsequent message discrimination developed through interaction with the attracted person. This process is derived mainly from online daters' internal desires, differing from the ought self pattern resulting from external forces \cite{filter_dating_2017}.

\subsubsection{Modes of Self-presentation.}

We relate our findings to Goffman’s impression management theory \cite{goffman1959presentation}, especially to the concepts of “front stage” and “backstage.” "Front stage" is where individuals perform to meet external societal expectations as performers perform to an audience they don't know. "Backstage" refers to the place where individuals expose their authentic selves without any audience. Our findings suggest that online dating users may also experience the switch between “front stage” and “backstage” modes when creating their dating profiles. Many participants indicated that they would actively modify their profiles to meet different external societal expectations, thus performing on the “front stage”. Occasionally, they would also seek authentic self-expression in what could be considered an online "backstage" setting.

Previous research suggested that self-presentation predominantly occurs in a background mode, shifting to a more active foreground mode only under important events \cite{schlenker1996impression}. However, our study found that self-presentation often took place on the “front stage.” One potential hypothesis is that online dating is inherently a special event requiring strategic self-marketing \cite{toma_separating_2008}. Participants emphasized the importance of presenting an appealing profile to attract attention, much as like in attention-driven marketplaces \cite{heino_relationshopping_2010}.

In addition to the "front stage" and "backstage" concepts, our findings suggest that there might be a third self-presentational strategy. Participants intended to remain “backstage” but imperceptibly appear “front stage” when asked to respond, as if it's an innate need to present themselves to others better than they actually are. This often resulted in profile misrepresentation due to expectations for themselves rather than external pressures. In this process, they may not be aware of misrepresentation due to self-perception bias, i.e., the "foggy mirror" phenomenon \cite{ellison_managing_2006}. In other words, they thought they were still “backstage” but actually were already on the “front stage,” where they were their own “audience”.

Another potential suggestion is that self-presentation in online dating is not just a dichotomy of "front stage" and "backstage" but a dynamic process influenced by the platform's features, user control, and algorithmic validation. Without being so literal about the "front stage" and "backstage" metaphor, Tong et al. \cite{tong_online_2016} discussed how online dating systems—whether see-and-screen, algorithmic, or blended—affect users’ decisions and self-presentation strategies. See-and-screen systems afford users significant personal control, while algorithmic systems provide validation through algorithmic matches. Blended systems offer the best of both worlds, enhancing perceived control due to external validation provided by algorithms while still retaining their ability to make their own decisions, thus boosting decision-making satisfaction and desire for relational pursuit. This may suggest that online daters also modify their profiles considering platform features and external algorithmic forces.


\subsubsection{Differences in Judging Others' Profiles vs. Users' Own Profiles.}


Moreover, we found that people may exhibit differences in judgments in how they perceive their own profiles versus their perception of others’ profiles. Many participants expressed their desire for others to see more than just their appearance by avoiding showing their faces in photos, while most of them also viewed physical attributes as the most important factor in assessing others’ profiles \cite{hitsch2010makes,toma2010looks}. Also, our participants used blurred or side-view photos to protect their own privacy, although they felt blurry photos in other’s profiles suggested falsehood. When looking for potential partners, our participants always quested for authenticity. Their own profiles, however, were often found to exaggerate or misrepresent to an acceptable extent, just as discovered by Hancock et al. \cite{hancock_truth_2007}.

This behavior seems to be contradictory or paradoxical at first glance, but upon closer examination, it can be understood as a reflection of the complex balance between privacy protection \cite{lo_contradictory_2013}, self-presentation within acceptable norms on the platform \cite{peng_be_2020}, and the quest for authenticity from others \cite{toma_what_2012}. In other words, this difference in judgment may be due to conflicting objectives. Users seem to carefully present themselves favorably without crossing into deception or violating community standards. 

The cautious self-presentation strategies and strict standards for others are likely related to self-protection in the online dating environment. Previous research has explored how the design of online dating apps can be improved by AI to reduce risks of harm for women on dating platforms, focusing on online conversations \cite{furlo_rethinking_2021} and face-to-face meetings \cite{RepurposingAI, WomenSafety, WomenSafety2}. Researchers also found that online dating users tend to use voice-based platforms to avoid being recognized through images, protecting their privacy and safety \cite{Shen_seeking_2024}. Our findings, however, highlight the need to design systems that support self-protection even through the first step of online dating, i.e., profile construction and evaluation.

\subsubsection{Curating Profile Features to Attract Other Users.}
In line with previous research \cite{gosnell2011self,toma_separating_2008}, our dating app users were also aware of the importance of utilizing different strategies in profile creation to attract others. In our study, users were found to attract potential dates through several photo selection and persona construction strategies, including stimulating imagination (i.e., making people want to explore more in this profile), first-photo selection, and constructing niche personas. One interesting finding is that people stimulate imagination by limiting facial display in photos, although physical attributes are considered the most important by many people. This finding supports previous work showing the importance of physical attributes in profiles \cite{whitty_revealing_2008}, and further suggests that the critical point of photo attractiveness may stand at provoking the desire to explore instead of only visually appealing. Also, this strategy can be counted as an instance of the incognito type of photos, which often represents the "playful nature of love" \cite{degen_profiling_2023}. Another important and novel finding is that people use visually captivating and humorous photos together to attract people at first sight, as well as reduce the distance brought by good-looking appearances. Interestingly, our participants mentioned that they would show niche and mysterious hobbies to attract attention, which is in accord with a recent finding showing that niche hobbies in profiles are becoming more popular and trump any other factors in attracting perfect mates \cite{noauthor_trainspotting_nodate}.

\subsubsection{Evaluating Others' Profiles through Photos and Text.}

In our study, the way participants evaluated others' profiles confirms many previous findings. For example, physical attributes were considered the most important and often built first impressions \cite{whitty_revealing_2008}, although overly perfect photos were deemed to be less authentic \cite{lo_contradictory_2013}. Our participants assessed text besides photos to gain a better understanding of the person and validate the authenticity of photos, which confirms previous findings indicating that the authenticity evaluation of photos is positively related to that of free-text components in profiles \cite{lo_contradictory_2013}. 

Moreover, small cues in profile photos or text played a significant role in the evaluation process \cite{roshchupkina_rules_2023}. For instance, P1 was attentive to photos showing animals, interpreting this as a sign of being a kind animal lover. Detailed and unique text attracted P2 and P15, suggesting that participants valued individuality in self-presentation. Participants noted that photos can convey more information and are more difficult to fake than text. However, they also recognized that text can evoke more subtle feelings that photos cannot provide. Our participants carefully evaluated others’ profiles through these small cues provided by both photos and text, looking for potential matches without crossing into deception.

\subsection{Design Implications}

Based on our analyses and summaries above, the dating software used by our participants is not limited to any specific country or sexual orientation. Our results are informative for the optimization of dating software as a whole. In this subsection, based on our findings, we discuss the implications for dating app design from the perspectives of presentation channels, matching tools, and trust-building.

\subsubsection{User Presentation Channels.}

In our study, participants crafted profiles with the \textit{ideal} or \textit{ought self} but sometimes still wanted to expose their true selves, which is in line with prior work suggesting that users engage in self-decorating behaviors but also desire their potential dating partners to accept their true selves \cite{seering_applications_2018}. Therefore, we believe that a two-stage user presentation would help online daters better present themselves. When first discovered by potential dating partners, only a superficial part of the profile, where users are allowed to choose what to include to show their best selves, should be displayed. As the relationship progresses, for instance, once a certain amount of chat time accumulates, further information can be revealed.

This intervention may help reduce the anxiety and inferiority complex users may feel during the initial presentation. Our participants, like P16, do not want to reveal too much personal information to strangers and may beautify their profiles and photos at the beginning, worrying about being judged for some aspects (e.g., appearance or height). However, they may want to present themselves as authentically as possible in the later stage. A previous study has explored how the ephemeral photo design for privacy concerns could influence match outcome, and found that people tended to share more personal information in the ephemeral photo than in the persistent photo, which led to better match outcome and receiver engagement \cite{he2020preserving}. Our approach may increase the number of matches by sharing more personal information gradually after first attracting others with an idealized image, providing a buffer and a process of mutual selection. 
This could be particularly beneficial for individuals with “invisible” disabilities \cite{porter_filtered_2017} and transgender dating platform users \cite{fernandez_i_2019}, as it enables them to establish a connection without the pressure of immediate disclosure while still having the choice to proactively disclose sensitive information in the later stage.

Furthermore, this method may help reduce the risk of privacy breaches while presenting the true self. As participants such as P4 mentioned, they are reluctant to share more personal photos (e.g., graduation pictures meant only for family and friends) on social platforms. As suggested by prior work, online daters should have the option to hide their personal information while having an information verification process to ensure safety and privacy \cite{lee_examining_2023}. The two-stage presentation allows users to protect these private details initially and reveal more only when they feel ready. It may also help reduce cumulative risks in online personal information disclosure \cite{nicol_revealing_2022}, as only limited information would be exposed in the first stage.

\subsubsection{More Precise Matchmaking Tools.}

Our findings suggest the need for more precise matchmaking tools that cater to users' preferences (e.g., personality traits, niche interests, etc.), enhancing the overall user experience with more detailed matching and increased accuracy \cite{tomita_matching_2022}. 

In online dating markets, users often want to pursue the most desirable partners with respect to some attributes (e.g., physical attributes) \cite{walster1966importance,hitsch2010matching,lee2008if}, while for other attributes (e.g., race or education), similar partners are wanted \cite{lewis2013limits,anderson2014political}. These two types of attributes are identified as competing and matching attributes \cite{bruch_aspirational_2018}. In our case, niche interests could be viewed as matching attributes. 

Previous research has suggested that identifying matching attributes from images, text, or self-reported data and making them searchable (i.e., findable via search) can produce a meaningful impact on match success \cite{ionescu_agent-based_2021}. This may potentially reduce the choice overload that can lead to lower decision-making satisfaction \cite{dangelo_there_2017}. Prior work testing the effect of anonymous browsing on various metrics has shown that anonymity resulted in fewer matches \cite{bapna2016one}. This may be because users who were gifted with anonymous browsing failed to leave weak signals through records of visits. This effect was particularly pronounced for women, as they tended to wait for their counterparts to make the first move and initiate the relationship. This study highlighted the importance of signals in initiating relationships. Moreover, signals of attraction-relevant traits were also suggested to be important in supporting decision-making in online conversations before face-to-face meetings \cite{supporting_zytko_2018,zytko_supporting_2020,facilitating_kim_2022}.

Our observation of user behavior reveals that online dating users typically match in two ways: as presenters, carefully crafting their profiles by incorporating signals into photos and text and conveying messages to potential matches; and as selectors, judging other users' profiles to see if their interests, hobbies, personalities match their preferences. The proposed matchmaking feature would identify these signals in profiles, and then make them searchable for potential partners. 

Therefore, more precise matchmaking tools can increase the number of matches by identifying matching attributes (i.e., niche hobbies) and making them searchable as suggested by prior work \cite{ionescu_agent-based_2021}. Meanwhile, it can help daters identify potential matches effectively through signaling matching attributes, balancing the number of choices to avoid overwhelming users \cite{dangelo_there_2017}.

\subsubsection{Establishing More Reliable Trust Relationships.}

As our data shows, mutual disclosure of information can enhance trust by establishing common ground \cite{lampe_familiar_2007}. This calls for a mechanism that can visualize potential cultural connections between users. For example, it can incorporate data from users' other social media accounts to show cities they have visited, restaurants they have reviewed, favorite animations, and other shared experiences.

Previous work has presented a veracity-enhancing design using existing social network data (i.e., Facebook) and significantly increased trust \cite{norcie2013bootstrapping}. Another study utilizing device-based location data found that this design could reduce transaction costs and establish common ground \cite{ma2017happens}. Previous work also suggests that paired collaborative activities could enhance evaluations of potential dates by developing trust \cite{zytko_enhancing_2015}. These interventions suggest the possibility of enhancing trust through common ground, such as location information, social connections, and shared experiences like restaurants or collaborative activities. This can potentially be realized through social media extensions on online dating platforms, or common shared spaces in online virtual platforms \cite{fu_i_2023}.

Current dating apps may have usability flaws that lead to an atmosphere of distrust and prejudice among users. Therefore, one of the main challenges facing online dating apps in the future is how to satisfy users' social needs while considering their demands for accuracy and authenticity. Through these improvements, dating apps can not only better meet users' needs but also build a healthier, more trustworthy communication environment on a deeper level.

\subsection{Limitations} \label{limitation}
This study, while providing insightful observations on online dating profile creation, is subject to certain limitations that must be considered. The 20 participants we interviewed, though diverse, may not capture the full spectrum of behaviors and attitudes prevalent in the broader online dating community. In particular, even though 7 of the participants had significant overseas experience, they were all college-educated young Chinese users. Western users may have less conservative ways of showing their profiles, for example, not including photos showing only part of their face or using photos with more skin exposure more frequently. The experiences and strategies described by our cohort thus may not fully represent the diverse range of user experiences across different demographics, which can be addressed by future studies.

Additionally, the reliance on self-reported data through semi-structured interviews introduces potential biases. Participants' responses could be influenced by social desirability bias, leading to either conscious or unconscious alteration of their experiences and strategies in online dating profile creation. This aspect raises concerns about the accuracy and generalizability of the findings. For example, participants may be too eager to show their effective attraction strategies in order to impress the researchers. Moreover, individuals may have a distorted self-perception to subconsciously make ideal profiles, which has not been extensively explored. Other methods, such as benignly observing user activities online or anonymous surveys, may complement our findings by providing a different mode of research probe. Another point to consider is, there is a possibility of bias introduced by the researchers themselves. The interaction between the interviewer and the participants may inadvertently influence the responses by the researchers’ expectations, demeanor, or even subtle cues.
In addition, the coding process may be influenced by researchers’ own perspectives, experiences, and cultural backgrounds, introducing another layer of potential bias.

Another notable limitation is the study's limited focus on the textual aspects of profile creation. This gap in our research arises from the primary emphasis on image selection strategies, potentially overlooking how textual content contributes to self-presentation and impression management in online dating profiles. The lack of in-depth exploration into textual self-presentation may lead to an incomplete understanding of the multifaceted nature of online dating profiles, where text and imagery work in tandem to convey a user's identity and interests. Additionally, this study lacks a detailed investigation of different dating platforms. The unique features (e.g., focusing on interests, emotional connections or academic backgrounds) of different dating platforms may influence self-presentation strategies.
Future studies could benefit from more specific and in-depth investigations tailored to the unique features and presentation styles of different platforms.

Furthermore, our study predominantly addresses the selection of images for online dating profiles, with minimal exploration into the realm of image editing. We focused on the choices users make from existing photographs rather than the alteration or enhancement of these images. The absence of an examination of image editing practices may overlook a significant aspect of self-presentation, especially in an online environment where editing images is a common practice for altering the perception of an individual's appearance and personality. Future work on image editing for profile creation may uncover additional insights that corroborate our findings in image selection or yield contradictory evidence that suggests newer ways of presenting attractiveness in dating profiles. Building on this, future studies should aim to explore the interplay between text and imagery in online dating profiles and examine the prevalence and impact of image editing on user perception and interaction.

\section{Conclusion}\label{sec:Conclusion}

Through 20 semi-structured interviews with online dating app users who are college-educated young adults in China, we investigated how online daters select their photos and text when crafting profiles, how they use dating profiles to present themselves, and how they evaluate others' profiles. Online daters strategically selected photos without identifying information to stimulate imagination or protect privacy. They presented visually appealing or humorous photos to target potential mates. We also found that participants exaggerated or misrepresented their traits in response to external and internal desires while striving to present their authentic selves. Our findings suggest that we should design systems that promote authentic self-expression while protecting privacy, provide more precise matchmaking tools, and help establish more reliable trust relationships.

\bibliographystyle{ACM-Reference-Format}
\bibliography{sample-base}


\newpage
\appendix

\section{Dating Platforms Used by Participants} \label{DatingApps}

\begin{table}[h!]
\centering
\begin{tabular}{lp{10cm}}
\toprule
\textbf{Apps} & \textbf{Description} \\
\midrule
Bumble & A social networking application. The profiles of potential matches are displayed to the user, who can "swipe left" to reject the candidate or "swipe right" to express interest. According to a June 2016 survey, 46.2 percent of its users are women. \\
Tantan & A popular Chinese dating app, similar to Tinder, designed for young people to meet new friends and potential dates through a swipe-based interface. People can view people around them and browse their photos. \\
Tinder &A globally used dating app, available in over 190 countries and in 56 languages. When two people swipe right on each other's profiles, they are matched based on their interests.\\
Soul & A Chinese social networking app focused on creating connections based on personality and interests rather than looks. Soul Test and Planet invite users to be divided into thirty planets through five quick questions. Different planets represent different personalities.\\
Summer & A social discovery app that encourages users to connect through small talk and common interests. A limited number of 25 messages allows users to make an initial assessment of matches.\\
Qing Teng Zhi Lian & A Chinese dating app for highly educated professionals with a focus on academic credentials, providing a platform for serious relationships and meaningful connections. \\
Hinge & It is a dating app that allows users to reject or attempt matches by responding to specific content in their profiles, encouraging users to find serious relationships through detailed profiles and engaging tips. The service emphasizes the uploading of various forms of user-generated content, such as photos, videos and "prompts", as a way of expressing personality and appearance.\\
Jimu & A Chinese social networking app that focuses on interest-based communities and activities to help users meet new friends and partners. \\
Blued & A social networking app tailored to the LGBTQ+ community, providing a platform for them to date, socialize and share content. It currently has over 40 million users in 193 countries. Its features include verified profiles, live streaming, timelines, and group conversations. \\
Fanka & A gay social app based on interest algorithms and geolocation in mainland China. Fanka is used in a similar way to Tinder, where users view personal information about recommended users, such as displayed avatars, nicknames, and photos, and then swipe left to indicate they are not interested, and right to indicate they are interested in making a match. It's a social app that connects users through shared hobbies and interests, fostering true friendship and community.\\
\bottomrule
\end{tabular}
\caption{Dating Platforms Used by Participants}
\label{tab:app_descriptions}
\end{table}

\clearpage

\section{Interview Questions} \label{interview}

\begin{itemize}
    \item \textbf{Profile Construction Process}
\end{itemize}

How do you choose photos for your profile? 

Are there any criteria?

Why do you choose this photo?

What do you want to show with this photo?

Do you consider the sequence of photos?

\begin{itemize}
    \item \textbf{Self-Presentation in Profiles}
\end{itemize}

What information do you mention in your dating app profile?

Is there any information you deliberately avoid mentioning?

Have you ever exaggerated or downplayed certain traits of yourself? Why?

How do you describe your personality and interests?

\begin{itemize}
    \item \textbf{Perception of Others' Profiles}
\end{itemize}

What do you first notice when browsing someone's profile?

What specific information or photos attract your attention?

What information or photos make you feel repulsed or uninterested?

How do you judge whether a profile is authentic or fake?

Have you ever decided not to pursue someone based on their information or photos?

\end{document}